\documentclass[%
 reprint,
 superscriptaddress,
preprintnumbers,
nofootinbib,
 amsmath,amssymb,
 aps,
]{revtex4-1}
\usepackage{amsmath}
\usepackage{amssymb}
\usepackage{amsthm}
\usepackage{MnSymbol}
\usepackage{nicefrac}
\usepackage{amsfonts}
\usepackage{dsfont}

\usepackage[markup=nocolor]{changes}

\usepackage{graphicx}

\usepackage[normalem]{ulem}

\usepackage{dcolumn}
\usepackage{bm}

\usepackage{color}
\usepackage[usenames,dvipsnames,svgnames,table]{xcolor}
\usepackage{slashed}
\usepackage{empheq}
\newcommand{\bea}{\begin{eqnarray}}
\newcommand{\eea}{\end{eqnarray}}
\newcommand{\be}{\begin{equation}}
\newcommand{\ee}{\end{equation}}

\begin{document}

\title{$d=4$ as the critical dimensionality of asymptotically safe interactions
}

\author{Astrid Eichhorn}
\email{eichhorn@sdu.dk}
\affiliation{CP3-Origins, University of Southern Denmark, Campusvej 55, DK-5230 Odense M, Denmark} 
\affiliation{Institut f\"ur Theoretische Physik, Universit\"at Heidelberg, Philosophenweg 16, 69120 Heidelberg, Germany}
\author{Marc Schiffer}
\email{m.schiffer@thphys.uni-heidelberg.de}
\affiliation{Institut f\"ur Theoretische Physik, Universit\"at Heidelberg, Philosophenweg 16, 69120 Heidelberg, Germany}

\begin{abstract}
We explore the question why our universe is four dimensional from an asymptotically safe vantage point. We find  hints that asymptotically safe quantum fluctuations of gravity can only solve the $U(1)$ Landau-pole problem in the Standard Model in four dimensions. This could single out the observed dimensionality of the universe as the critical dimensionality of asymptotically safe interactions.
\end{abstract}

\maketitle

\section{Motivation}
To the best of our knowledge, our universe is four-dimensional. Experiments reach down to the micrometer level with tests of the gravitational inverse-square law \cite{Hoyle:2004cw,Kapner:2006si,Adelberger:2009zz,Murata:2014nra}, conversely up to several TeV  with searches at the LHC \cite{Aaboud:2016ewt,Sirunyan:2018ipj}, see also \cite{PhysRevD.98.030001} for an overview, and find no indications for extra dimensions at these distances. This motivates us to ask: What is special about four dimensions? For instance, starting from a string-theoretic paradigm for the nature of the microscopic building blocks of our universe, $d=10$ appears as the critical dimension of the superstring. Here, we take a step towards discovering whether a quantum-field theoretic description of the fundamental building blocks of nature might also give rise to a critical dimensionality.\\
Specifically, we work within the asymptotic-safety framework. Asymptotic safety, first proposed for quantum gravity by Weinberg \cite{Weinberg:1980gg}, is a quantum field theoretic paradigm that could make a predictive description of quantum gravity possible. It is particularly appealing in its simplicity, as it is based on a generalization of the powerful principle of asymptotic freedom and describes the quantum properties of spacetime in terms of quantum fluctuations of the metric. Asymptotic safety corresponds to an interacting ultraviolet (UV) fixed point of the Renormalization Group (RG) flow,
such that the microscopic dynamics is characterized by scale-symmetry.
Indications for the existence of the Reuter fixed point were discovered in Euclidean gravity 
\cite{Reuter:1996cp,Reuter:2001ag, Lauscher:2001ya,Litim:2003vp,Codello:2008vh, Benedetti:2009rx, Manrique:2011jc,Christiansen:2014raa,Becker:2014qya, Demmel:2015oqa, Gies:2016con, Denz:2016qks, Christiansen:2017bsy, Falls:2018ylp} and Euclidean  gravity-matter systems 
\cite{Narain:2009fy,Dona:2013qba,Meibohm:2015twa,Dona:2015tnf,Eichhorn:2016vvy,Christiansen:2017cxa,Biemans:2017zca,Alkofer:2018fxj,Eichhorn:2018akn,Eichhorn:2018ydy,Eichhorn:2018nda}, including in higher dimensions \cite{Litim:2003vp,Fischer:2006at,Fischer:2006fz,Falls:2015cta,Dona:2013qba}. For recent reviews see, e.g., \cite{Reuter:2012id, Ashtekar:2014kba,Eichhorn:2017egq, Percacci:2017fkn,Eichhorn:2018yfc,Wetterich:2019qzx}. Asymptotic safety is not unique to quantum gravity; interacting  RG fixed points exist in quantum field theories with fermions, scalars as well as in gauge theories across different dimensions, see, e.g., \cite{Peskin:1980ay,Gies:2003ic,Codello:2008qq,Braun:2010tt}, including gauge-Yukawa systems in four dimensions \cite{Litim:2014uca,Litim:2015iea}.

In quantum field theory, $d=4$ is a special dimension, as it is the critical dimension for the interactions in the Standard Model, resulting in perturbative renormalizability.
On the other hand, the critical dimension for Einstein gravity is $d=2$, as this is the dimensionality in which the Newton coupling is dimensionless. 
Combining Standard Model matter and gravity within the asymptotic safety paradigm thus raises the intriguing question whether the combined model gives rise to a preferred dimensionality and what its value might be.

\section{Asymptotically safe quantum gravity and the U(1) gauge sector of the Standard Model}
In this letter, we focus on the Abelian gauge sector of the Standard Model, featuring the scale-dependent hypercharge coupling $g_{\scriptscriptstyle Y}(k)$, where $k$ is a momentum scale. Quantum fluctuations of charged matter effectively turn the vacuum into an antiscreening medium. Accordingly, the value of the gauge coupling $g_{\scriptscriptstyle Y}(k)$ decreases as the momentum scale $k$ is lowered. This is encoded in the beta function which reads
\begin{equation}
	k\partial_kg_{\scriptscriptstyle Y}(k)=\beta_{g_{\scriptscriptstyle Y}}\big|_{\rm SM}=\frac{1}{16\pi^2}\frac{41}{6}g_{\scriptscriptstyle Y}^3,
\end{equation}
at one-loop order.
Hence, the scale-dependence of the gauge coupling is given by
\begin{equation}
g_{\scriptscriptstyle Y}^2(k)=\frac{g^2_{\scriptscriptstyle Y}(k_0)}{1-\frac{1}{8\pi^2}\frac{41}{6}g^2_{\scriptscriptstyle Y}(k_0)\ln\left(\frac{k}{k_0}\right)},
\end{equation}
such that a finite value of $g_{\scriptscriptstyle Y}$ at $k_0$ is linked to a divergence in the coupling in the UV, the so-called Landau pole. The Landau pole suggests a breakdown of the Standard Model at a finite scale \cite{GellMann:1954fq,Gockeler:1997dn,Gockeler:1997kt,Gies:2004hy}. An ultraviolet completion requires the presence of new physics. It is intriguing that a minimalistic extension by asymptotically safe quantized gravitational fluctuations without any additional degrees of freedom could suffice to induce a predictive ultraviolet completion \cite{Harst:2011zx,Eichhorn:2017lry,Eichhorn:2018whv}:
Asymptotically safe quantum gravity results in an additional contribution to the flow of the gauge coupling. It is leading order in the gauge coupling, as it only enters linearly in $g_{Y}$. Including the gravitational contribution, the beta-function for $g_{Y}$ in an asymptotically safe setting is given by
\begin{equation}
\beta_{g_{\scriptscriptstyle Y}}=-f_g g_{\scriptscriptstyle Y}+\frac{1}{16\pi^2}\frac{41}{6}g_{\scriptscriptstyle Y}^3 +\mathcal{O}(g_{\scriptscriptstyle Y})^4.
\label{eq: betag1}
\end{equation}  
The gravitational term dominates at small couplings. Explicit calculations using functional Renormalization Group techniques yield $f_g\geq0$ \cite{Daum:2009dn,Folkerts:2011jz,Harst:2011zx,Christiansen:2017gtg,Eichhorn:2017lry,Christiansen:2017cxa}, indicating an antiscreening effect of metric fluctuations.
 An intuitive way to understand the gravitational contribution is through an effective dimensional reduction: In $d<4$, the beta function of the gauge coupling features a similar term due to the canonical dimension of the gauge coupling, $[g_{\scriptscriptstyle Y}]=2-d/2$. Hence, a gauge field on a fluctuating instead of a fixed background ``sees" an effectively reduced dimensionality.

More specifically, $f_g$ is a function of the gravitational couplings, and proportional to the dimensionless Newton coupling measured in units of $k$, $g(k)= G_N(k)k^2$. We make the assumption that below the Planck scale, gravity is classical. In this regime, $G_N$ exhibits essentially no scale dependence, such that $g(k)\approx0$ for $k^2<M_{\rm Pl}^2\sim G_N^{-2}$ and the scale dependence of the Abelian gauge coupling is completely determined by quantum fluctuations of Standard Model fields. At and above $k^2\approx M_{\rm Pl}^2$, quantum fluctuations of gravity are important. They generate a scale-invariant fixed-point regime, i.e., $g(k)=g_*=const$ for $k^2\geq M_{\rm Pl}^2$. As a consequence, $f_g=const$ holds in this regime. 

At sufficiently small $g_{\scriptscriptstyle Y}(k\geq M_{\rm Pl})$, the linear term in Eq.~\eqref{eq: betag1} dominates. Due to the antiscreening nature of asymptotically safe quantum fluctuations of gravity, the Abelian gauge coupling becomes asymptotically free. Accordingly, gravity fluctuations render this sector ultraviolet complete. A second, interacting fixed point provides another possibility for an ultraviolet completion that even results in a retrodiction of the gauge  coupling \cite{Harst:2011zx,Eichhorn:2017lry,Eichhorn:2018whv,Eichhorn:2017muy}.

A similar contribution as in Eq.~\eqref{eq: betag1} has been discussed in a different setting, namely in perturbation theory, see, e.g., \cite{Robinson:2005fj,Pietrykowski:2006xy,Toms:2010vy,Anber:2010uj}, where no scale-dependence of the gravitational interactions was taken into account. In the absence of a cosmological constant \cite{Toms:2009vd}, a field-redefinition allows to shift the gravitational effect in perturbation theory to higher-order interactions \cite{Ellis:2010rw}. We stress the difference of this setting to a common asymptotically safe fixed point for gravity and matter which generically features a finite cosmological constant. Field redefinitions cannot change the universal properties of such a fixed point, such as the critical exponents, which include $f_g$ in the present case. In a different choice of coordinates in field space, the universal physical consequences of a fixed point are therefore expected to be encoded in a more involved fashion.
\begin{figure}[!t]
	\centering
\includegraphics[width=\linewidth]{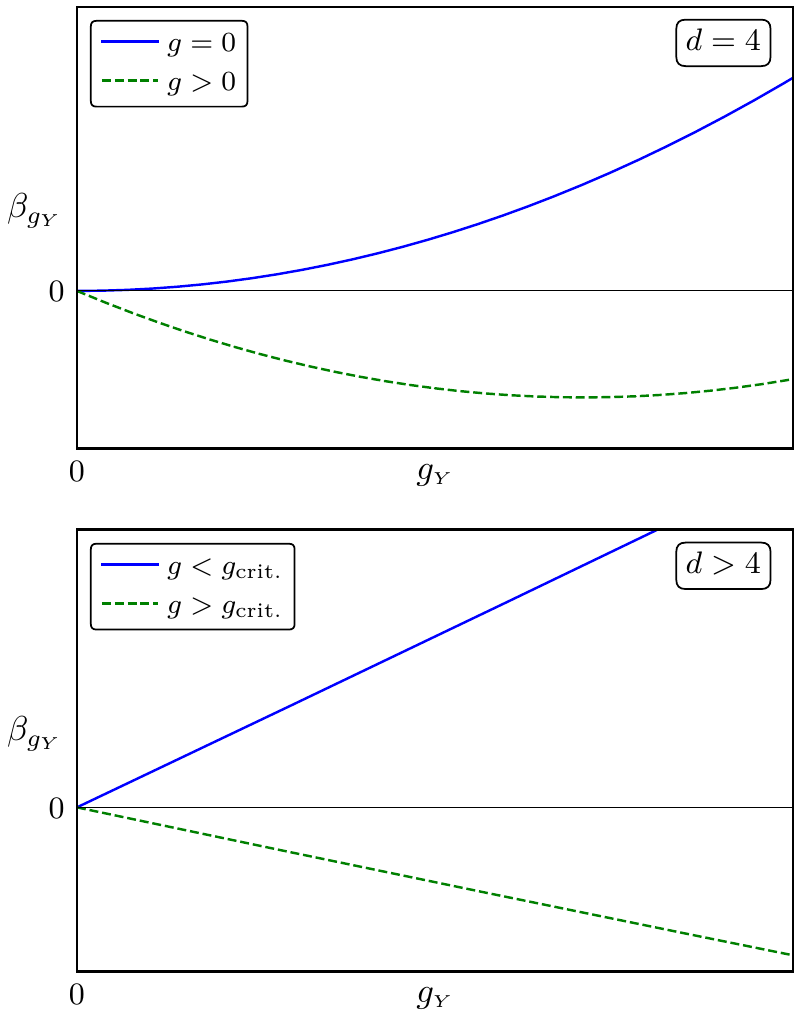}
	\caption{\label{fig: beta} 
	Upper panel: $\beta_{g_{\scriptscriptstyle Y}}$ in $d=4$ shows an infrared attractive fixed point at $g_{\scriptscriptstyle Y}=0$ without gravity, resulting in a Landau pole.  With gravity, the coupling becomes asymptotically free.
	Lower panel: To solve the Landau-pole problem in $d>4$, the gravitational coupling strength must be larger than a non-zero critical value $g_{\rm crit}$ in order to induce asymptotic freedom.}
\end{figure}

\section{No asymptotically safe UV completion in $d>4$}

We next turn our attention to the case $d>4$. We make the assumption that the extra dimensions are compactified, such that observational bounds on the dimensionality of our universe at large scales can be met. Further, we focus on the asymptotically safe fixed-point regime in the far ultraviolet, at distance scales much smaller than any compactification scale.

The major difference between $d=4$ and $d>4$ is the canonical mass dimension of the gauge coupling, $[g_{\scriptscriptstyle Y}]=2-d/2$. The coupling is marginal in $d=4$, but canonically irrelevant in $d>4$, exacerbating the Landau-pole problem without gravity. Therefore, its beta-function reads
\begin{equation}
\beta_{g_{\scriptscriptstyle Y}}=g_{\scriptscriptstyle Y}\left(\frac{d-4}{2}-f_g(d)\right)+\mathcal{O}(g_{\scriptscriptstyle Y}^3).
\end{equation}
In fact, as we will show, $f_g(d)\,>\, 0$ for all $d\geq4$. Thus gravity always acts like an effective \emph{decreased} dimensionality. If the ``effective dimensionality" is smaller than 4, then the gauge coupling is asymptotically free. Accordingly, $f_g(d=4)>0$ is sufficient to render the gauge coupling asymptotically free in $d=4$, as highlighted in the upper panel of Fig.~\ref{fig: beta}. In $d>4$ spacetime dimensions, the gravity-induced term has to compete with an actual canonical-dimension term. The latter counteracts  the gravitational term. To reduce the ``effective dimensionality" sufficiently, $f_g(d)$ has to exceed the critical value
\begin{equation}
f_{g,\,\rm crit}(d)=\frac{d-4}{2}.
\end{equation}
In our setting, $f_g>f_{g,\,\rm crit}$ is a necessary condition to avoid the Landau pole and render the Abelian gauge sector UV-complete without the need for new physics beyond gravity, cf.~the lower panel of  Fig.~\ref{fig: beta}.
Unless the strength of the gravitational contribution grows sufficiently to ensure $f_g(d)>f_{g, \, \rm crit}(d)$ as the dimensionality of spacetime is increased, the gravitational solution to the Abelian triviality problem could only be available in $d=4$.

As an increase in the dimension leads to an increased number of propagating gravitational degrees of freedom with $d(d-3)/2$, one might expect $f_g(d)$ to grow as a function of dimensionality.
Yet, the impact of quantum fluctuations also depends on the $d$-dimensional integral over the momentum of virtual field configurations. This is related to the volume of the $d$-dimensional unit sphere and actually goes to zero with  $\Gamma[d/2]^{-1}$ as $d \rightarrow \infty$.
Therefore, for fixed values of the gravitational couplings, $f_g(d)$ is expected to decrease with increasing $d$.
This is the opposite behavior to that required to satisfy the condition $f_g(d)>f_{g, \,\rm crit} = (d-4)/2$. This could be circumvented by an appropriate dependence of the asymptotically safe fixed-point value of the Newton coupling $g_*$ on the dimensionality. In fact, $f_g$ is proportional to $g_*(d)$, and thus depends on $d$ also indirectly, $f_g(d)= f_g(d, g_*(d))$. Accordingly, if gravity becomes increasingly non-perturbative, such that $g_*(d)$ increases, then $f_g(d,g_*(d))>f_{g,\, \rm crit}(d)$ could be possible, even though $f_g(d, g={\rm const})$ decreases.
Therefore, in $d>4$, a gravitational solution to the Landau-pole problem requires an increasingly non-perturbative nature of gravity  in order for the fixed-point value of the Newton coupling to grow sufficiently fast to compensate for both effects.

However, a weakly coupled nature of asymptotically safe quantum gravity appears favored due to several arguments. \\Firstly, the Standard Model couplings are perturbative at the Planck scale. This suggests that the underlying UV completion, which generates these values, is (near-) perturbative in nature. In fact, hints for this scenario have been discovered in $d=4$ \cite{Shaposhnikov:2009pv,Eichhorn:2017ylw,Eichhorn:2017lry,Eichhorn:2017als,Eichhorn:2018whv}. \\
Secondly, a near-perturbative description of gravity is favored from a computational point of view due to its controllability.
Indeed, in $d=4$, indications for a near-perturbative, i.e., near-Gaussian behavior of asymptotically safe gravity have been found in \cite{Falls:2013bv,Falls:2014tra,Falls:2017lst,Eichhorn:2018akn,Eichhorn:2018ydy,Falls:2018ylp,Eichhorn:2018nda}.\\
Thirdly, the mechanism to induce asymptotic safety in gravity appears to be a competition of quantum and canonical scaling for the Newton coupling. This mechanism is familiar from non-gravitational models \cite{Wilson:1971dc,Peskin:1980ay,Gies:2003ic,Braun:2010tt} and allows to connect interacting fixed points in higher dimensions to the free fixed point in the critical dimension of the model. Such a connection allows to recover asymptotically safe fixed points from appropriate Pad\'e-resummations of the $\epsilon$-expansion that can be calculated perturbatively. \\
Fourth, one-loop perturbation theory in gravity actually provides explicit hints for an asymptotically safe fixed point \cite{Niedermaier:2009zz,Niedermaier:2010zz}.

Moreover, there are indications that the strength of gravitational fluctuations must not be too large to allow for a UV complete model \cite{Eichhorn:2012va,Eichhorn:2016esv,Christiansen:2017gtg,Eichhorn:2017eht} as, beyond the weak-gravity regime, quantum gravity triggers new divergences in the matter sector. Specifically, the interacting nature of asymptotically safe gravity percolates into the matter sector and induces higher-order interactions in the ultraviolet \cite{Eichhorn:2011pc,Eichhorn:2012va,Meibohm:2016mkp,Eichhorn:2016esv,Eichhorn:2017eht,Eichhorn:2017sok,Christiansen:2017gtg,Eichhorn:2018nda}.
At small values of the gravitational coupling strength, these higher-order operators are benign: The ultraviolet fixed-point values of these couplings are small, and they remain irrelevant even under the impact of quantum gravity.
At larger values of the  gravitational coupling strength, the formal fixed-point solution for some of the higher-order operators is complex. This is not physically viable 
and restricts the allowed parameter space for the microscopic gravitational couplings to the weak-gravity regime.

More specifically, asymptotically safe gravity generates a term quartic in the $U(1)$ field strength of the form $w_2 \left(F^2\right)^2$ \cite{Christiansen:2017gtg}. At sufficiently large gravitational coupling  a novel UV-divergence is triggered by purely gravitational contributions to the flow of $w_2$. These generate a non-vanishing flow even at $w_2=0$ by contributions to the flow that are proportional to the Newton coupling $g$, but independent of the coupling $w_2$ itself. Schematically, the flow of $w_2$ takes the form of
\begin{equation}
\beta_{w_2}=A_0(g)+w_2\, A_1(g)+w_2^2A_{2},
\label{eq: w2beta}
\end{equation}
where $A_0$ and $A_1$ are functions of $g$, such that $A_0\to0$ for $g\to 0$.
Thus, at vanishing $g$, i.e., in the absence of 
gravity, $\beta_{w_2}=0$ is solved by $w_{2*}=0$. This is no longer the case at finite $g$, where
\be
w_{2*}= \frac{-A_1(g)}{2A_2} \pm \sqrt{\frac{A_1(g)^2}{4 A_2^2}- \frac{A_0(g)}{A_2}}.
\ee
Thus, $w_2$ cannot be set to zero in the presence of gravity, and instead features a shifted Gaussian fixed point proportional to the Newton coupling, $w_{2*}\sim g$, for small $g$. Further, it turns out that the shifted Gaussian fixed point vanishes into the complex plane at 
\begin{equation}
A_{0,\, \rm crit}(g)= \frac{A_1^2(g)}{4A_2}, 
\end{equation}
which can be re-expressed as a bound on $g$.
Beyond this critical strength of the gravitational interaction, the novel divergences in $w_2$ associated to the lack of an ultraviolet fixed point signify the UV-incompleteness of the model.

Accordingly, a viable UV completion has to simultaneously fulfil the conditions
\begin{equation}
f_g>f_{g,\rm crit}\quad\text{and}\quad 
A_0 \leq A_{0,\,\rm crit}.
\end{equation}
Both conditions can be re-expressed as conditions on the microscopic values of the gravitational couplings. While the former requires a sufficiently large $g_*$, the latter necessitates a sufficiently small $g_*$. Whether these opposing requirements can be reconciled requires an explicit calculation.

We calculate $f_g(d)$ and $A_0,\,A_1$ and $A_2$
with functional RG techniques \cite{Wetterich:1992yh, Ellwanger:1993mw,Morris:1993qb}. The functional RG is based on the scale dependent effective action $\Gamma_k$ and realizes the Wilsonian idea of momentum-shell wise integration of quantum fluctuations. The Wetterich equation encodes the scale dependence of the dynamics and is given by
\begin{equation}
\partial_t\Gamma_k=\frac{1}{2}\rm Tr\left[\left(\Gamma^{(2)}_k+R_k\right)^{-1}\partial_tR_k\right],\label{eq:Wetteq}
\end{equation}
where $\partial_t=k\partial_k$. Here, $\Gamma^{(2)}_k$ is the second functional derivative of the scale-dependent effective action. The infrared regulator $R_k$ suppresses low-energy modes in the generating functional underlying $\Gamma_k$. At the same time, it ensures UV  finiteness and implements the step-wise integration of quantum fluctuations according to their momentum in the flow equation \eqref{eq:Wetteq}. We choose a Litim type cutoff \cite{Litim:2001up}
\begin{equation}
R_k=Z_{\Phi}(k^2-p^2)\Theta(k^2-p^2),
\end{equation}
where $Z_{\Phi}$ stands for the wave-function renormalization of the respective field. For introductions and general reviews of the method, see, e.g.,
\cite{Berges:2000ew, Delamotte:2007pf, Rosten:2010vm, Braun:2011pp};
specifically for the application in gravity and gauge theories, see, e.g., \cite{Pawlowski:2005xe, Gies:2006wv,Reuter:2012id}.
The flow equation has a diagrammatic representation in terms of one-loop diagrams with a regulator insertion. Due to their dependence on the dressed propagator they can be understood as resummations of perturbative contributions.

\begin{figure}[!t]
	\centering
	\includegraphics[width=0.9\linewidth]{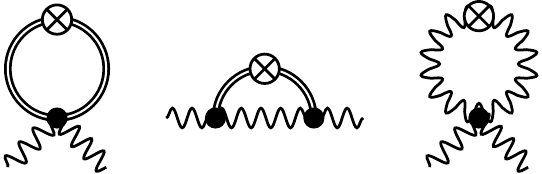}
	\caption{\label{fig:diageta} Diagrams contributing to the anomalous dimension $\eta_A$. Metric fluctuations are denoted by straight double lines and gauge bosons by wiggly lines. The regulator insertion $\partial_t R_k$ is denoted by a crossed circle. For the two-vertex diagram, only one representative is shown.}
\end{figure}

The direct gravitational contribution to $\beta_{g_{\scriptscriptstyle Y}}$ is encoded in the first two diagrams in Fig.~\ref{fig:diageta}, such that
\begin{equation}
f_g =-\frac{\eta_A\Big|_{\rm grav}}{2}.
\end{equation}

\begin{figure}[!t]
	\centering
	\includegraphics[width=\linewidth]{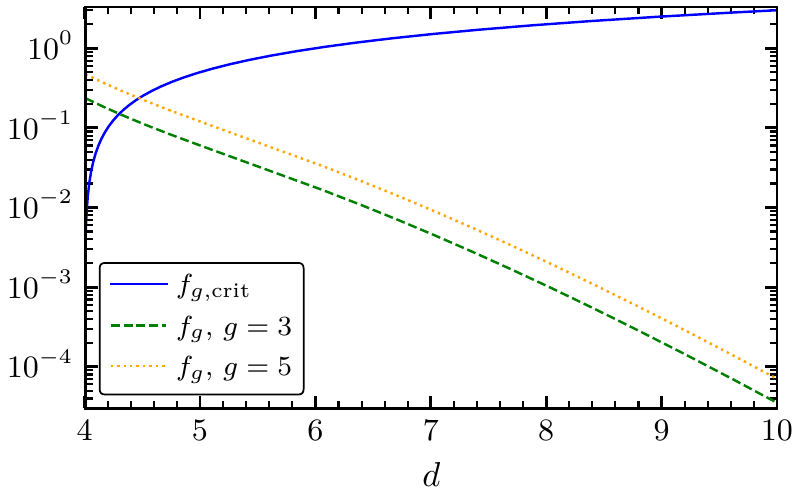}
	\caption{\label{fig:fgcrit} 
	We show how the values of $f_g$, $f_{g,\rm crit}$  exhibit a contrary behavior as a function of the 
	dimensionality. For purposes of illustration we choose $\lambda=0$ and $w_2=0$ and show $f_g$ for $g=3$ and $g=5$.
}
\end{figure}

We work within a truncation of the full dynamics given by
\begin{align}
\Gamma_k=
&\frac{Z_A}{4}\int \!\! \mathrm{d}^{d} x \sqrt{g}g^{\mu\rho}g^{\nu\kappa}F_{\mu\nu}F_{\rho\kappa}\notag\\
&+\frac{w_2k^{-d}}{8}\int \!\! \mathrm{d}^{d} x \sqrt{g}(g^{\mu\nu}g^{\rho\lambda}F_{\mu\lambda}F_{\nu\rho})^2\notag\\
&-\frac{1}{16\pi\, g\,k^{-d+2}} \int \!\! \mathrm{d}^{d} x \sqrt{g} \, (R - 2 \lambda k^2)+S_{\rm gf.},
\end{align}
where we have already expressed all couplings by their dimensionless counterparts in units of $k$. The gravitational dynamics is encoded in the values of the two couplings $g$ and $\lambda$.
For the gauge-fixing, we use a standard covariant gauge condition in the Landau-gauge limit for the photons and metric fluctuations. 
To evaluate the RG flow, we use the Mathematica package \emph{xAct} \cite{Brizuela:2008ra,Brizuela:2008ra,2008CoPhC.179..597M,2007CoPhC.177..640M,2008CoPhC.179..586M} as well as the FORM-tracer \cite{Cyrol:2016zqb}.

\begin{figure}[!t]
	\centering
	\includegraphics[width=.9\linewidth]{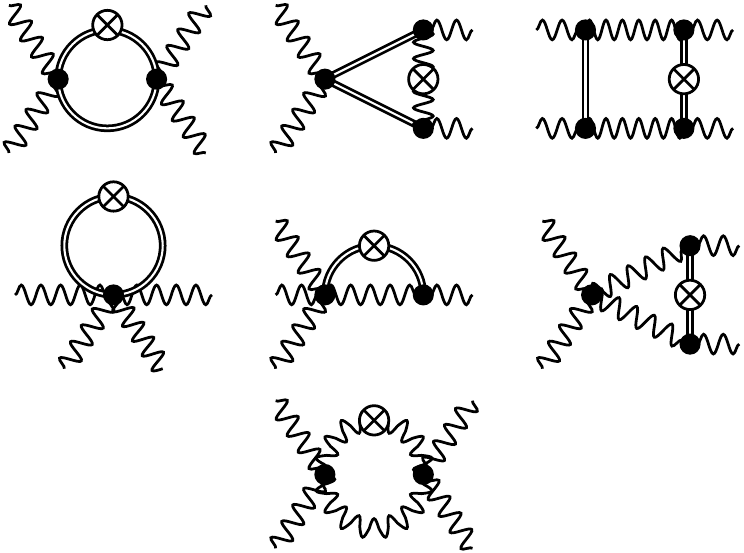}
	\caption{\label{fig:diagr dw2} Diagrams contributing to the flow of $w_2$. The regulator insertion is understood to appear on each of the internal propagators and only one representative of each class of diagrams is shown. The first line of diagrams are the induced contributions making up $A_0$, cf.~Eq.\eqref{eq: w2beta}.}
\end{figure}

Our results are given by
\begin{align}
f_g(d)
=&
g\frac{2^{1-d}\pi^{1-\tfrac{d}{2}}\left(16+(d-2)d\left(12+(d-9)d\right)\right)}{(d-2)d\, \Gamma\!\left[2+\frac{d}{2}\right](1-2\lambda)^2}\left(2+d\right)\notag\\
&
+
g\frac{2^{3-d}((d-2)d-2)\pi^{1-\tfrac{d}{2}}}{(d-2)\Gamma\!\left[3+\frac{d}{2}\right](1-2\lambda)}\left((4+d)
+
\frac{(4+d)}{1-2\lambda}\right)\notag\\
&-w_{2*}(4+d)\frac{4+d(d-1)}{2^{d+1}\pi^{\tfrac{d}{2}}\Gamma\!\left[3+\frac{d}{2}\right]},
\end{align}
where the contribution in the last line corresponds to the third diagram in Fig.~\ref{fig:diageta}. It depends on $g$ implicitly, as $w_{2*}$ is a function of $g$, determined by solving $\beta_{w_2}=0.$ Our first key result consists in the comparison of $f_g(d)$ to $f_{g, \, \rm crit}(d)$, cf.~Fig.~\ref{fig:fgcrit}. As expected, these exhibit opposite behaviors with increasing $d$. Accordingly, a gravitational solution to the Landau-pole problem becomes more difficult to achieve as $d$ increases. In fact, a strong increase of the microscopic gravitational coupling $g$ is the only possibility to compensate for the decline of $f_g(d)$ as a function of $d$.
To highlight that the required increase of the microscopic gravitational coupling appears irreconcilable with a well-controlled UV behavior of the entire Abelian sector, we evaluate $w_{2\, \ast}$ by studying its beta function
\begin{align}
\beta_{w_2}=&(d+2\eta_A)w_2\notag\\
+&g^2\bigg(\!\frac{2^{7-d}\pi^{2-\tfrac{d}{2}}}{(d-2)^2(1-2\lambda)^2\Gamma\!\left[3+\tfrac{d}{2}\right]}\bigg(\!(4\!-\!d)(4\!-\!10d\!+\!d^3)\notag\\
&+\!\frac{d^7\!+\!d^6\!-\!30d^5\!+\!36d^4\!+\!136d^3\!-\!544d^2\!+\!1152d\!-\!512}{2^3d(1-2\lambda)}\bigg)\bigg)\notag\\
+&g\,w_2\bigg(\!\frac{2^{8-d}\pi^{1-\tfrac{d}{2}}}{(d-2)(1-2\lambda)\Gamma\!\left[3+\tfrac{d}{2}\right]}\bigg(\!(3+(d-5))\notag\\
&+\frac{d^6-13d^5-48d^4+276d^3+112d^2-256d-256}{2^7d(1-2\lambda)}\bigg)\bigg)\notag\\
+&w_2^2\frac{2^{1-d}(48+d(6+d+d^2))}{\pi^{d/2}\Gamma\!\left[3+\tfrac{d}{2}\right]}.
\end{align}
The terms proportional to $g^2$, $g\,w_2$ and $w_2^2$ correspond to the diagrams of the first, second and third row in  Fig.~\ref{fig:diagr dw2}, respectively.
In the above expressions we work in the approximation where anomalous dimensions arising from the regulator-insertion are neglected.

\begin{figure}[!t]
	\centering
	\includegraphics[width=\linewidth]{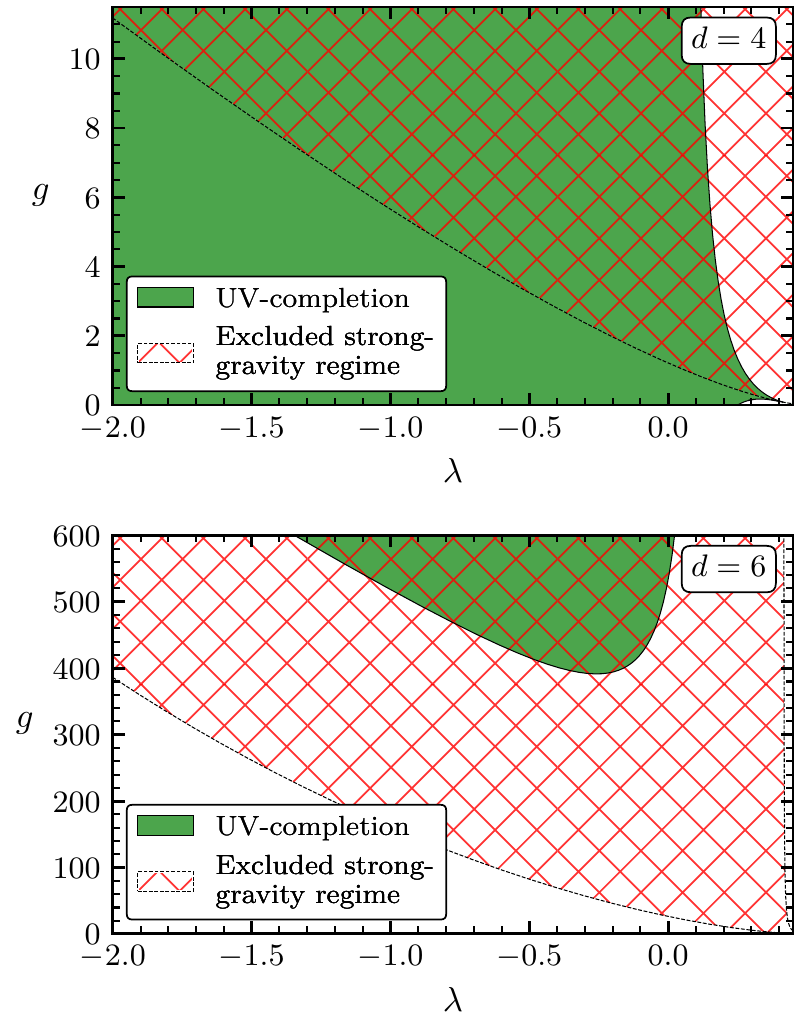}
	\caption{\label{fig:46d} We indicate the region where $f_{g}>f_{g,\, \rm crit}$ in green and the region where $A_0>A_{0,\, \rm crit}$ in red hatched in $d=4$ (upper panel) and $d=6$ (lower panel). An ultraviolet completion can only be achieved in that part of the green region that does not overlap with the red region.
	}
\end{figure}

We can now study explicitly, whether the two conditions $f_{g}(d)>f_{g,\, \rm crit}(d)$ and $A_0(g)\leq A_{0, \, \rm crit}(g)$ can be met simultaneously for any values of the gravitational couplings $g$ and $\lambda$, as the dimensionality is increased.
The green area of the upper panel in Fig.~\ref{fig:46d} indicates where the condition $f_{g}>f_{g, \, \rm crit}$ holds in $d=4$. 
The red hatched area indicates where strong gravity fluctuations induce new divergences in the matter sector, i.e., where $A_0>A_{0, \,\rm crit}$, such that the fixed point of $w_2$ lies off the real axis. Therefore, the overlap of the red hatched region with the green region is removed from the allowed gravitational parameter space. In $d=4$, the fixed-point values for gravity in the presence of minimally coupled, non-interacting Standard Model matter actually fall into the remaining green region \cite{Dona:2013qba,Biemans:2017zca,Christiansen:2017bsy,Alkofer:2018fxj}. This indicates that quantum gravity could solve the Landau-pole problem in $d=4$.

\begin{figure}[!t]
	\centering
	\includegraphics[width=\linewidth]{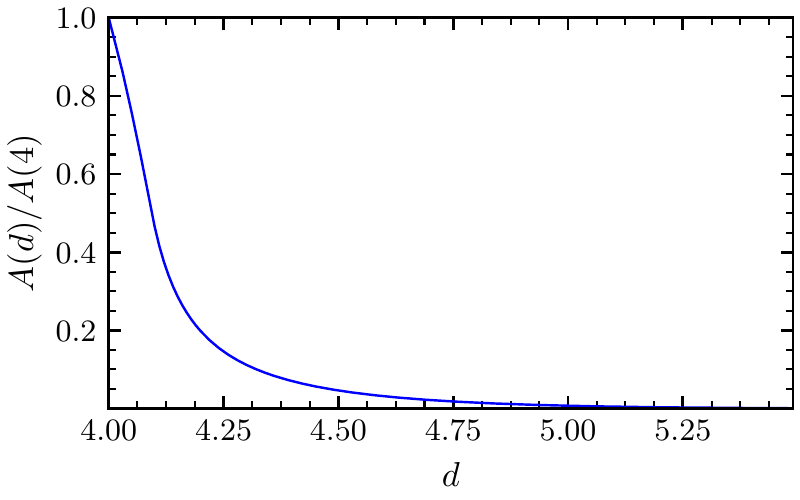}
	\caption{\label{fig:areas} Area where a UV completion is possible for $g\in(0,1000),\lambda\in(-1500,0.5)$ as function of the spacetime dimension.}
\end{figure}

The situation is very different in $d>4$: We study the area of the allowed region in a very extended range of gravitational couplings, $g\in(0,1000)$ and $\lambda\in(-1500,0.5)$, and find that it is already very small in $d=5$ and shrinks to zero at $d \approx 5.8$, as the green region overlaps completely with the red hatched region, cf.~Fig.~\ref{fig:areas}. Already in $d=6$, the green region has moved very far into the red hatched region, cf.~lower panel in Fig.~\ref{fig:46d}. Accordingly, in our approximation in $d=6$  and above the Abelian gauge sector remains UV-incomplete even in the presence of gravity irrespective of which fixed-point values $g_{\ast}$ and $\lambda_{\ast}$ are realized.

Our explicit calculation leading to Fig.~\ref{fig:46d} and \ref{fig:areas} is subject to systematic errors due to our use of a truncation for the calculation of the RG flow. Extended truncations will lead to changes in the fixed-point values for $g, \lambda$, as well as deformations of the boundaries shown in Fig.~\ref{fig:46d}. We highlight that very large deformations would be required to change our conclusion that there is no viable area in the gravitational coupling space, cf.~Fig.~\ref{fig:areas} within the explored range $g\in(0,1000), \, \lambda \in (-1500,0.5)$ in $d \geq 6$.  Further, the qualitative aspects of the scenario put forward here are unaffected by the choice of truncation, and single out $d=4$ as the only dimensionality in which a gravitational solution to the Landau-pole problem appears achievable.

\section{Summary}
A simple description of the microscopic building blocks of nature in terms of the quantum fields of the Standard Model, together with quantum-gravity effects encoded in metric fluctuations, appears to be compatible with current data from the LHC. In such a setting, scale-invariance takes center stage at (trans)Planckian energies. Most intriguingly, this principle appears to potentially be powerful enough to restrict free parameters of the Standard Model, such as the low-energy values of several couplings \cite{Shaposhnikov:2009pv,Harst:2011zx,Eichhorn:2017ylw,Eichhorn:2017lry,Eichhorn:2018whv}. 
A similar point of view has been advocated in \cite{tHooft:2016uxd}. Yet, free parameters, connected to the quantum nature of spacetime, remain. We have tackled one of them, namely the question why we live in four spacetime dimensions. We have found indications that $d=4$, and possibly $d=5$, is the only dimensionality in which two competing effects can be reconciled in our setting: On the one hand, a gravitational solution to the Landau-pole problem requires gravitational fluctuations to be strong enough. Only then can they lower the effective dimensionality of spacetime, defined from the canonical plus the gravitational quantum scaling of the gauge coupling, to below four. Yet, the canonical contribution increases as a function of the spacetime dimensionality, requiring the gravitational contribution to increase with it. As our first key result, we have shown that the gravitational contribution actually exhibits the opposite behavior and decreases with $d$ due to a loop-suppression if the gravitational coupling is held fixed. Thus, a gravitational solution to the Landau-pole problem necessitates a strong growth of the microscopic gravitational coupling $g_*$.
On the other hand, gravitational fluctuations should remain near-perturbative in the ultraviolet. One explicit manifestation of this requirement is the existence of new divergences in the matter sector that appear once gravity fluctuations become strong. These divergences appear in otherwise benign higher-order couplings, and prevent the viability of a gravity-induced ultraviolet completion. As a consequence, there is an excluded ``strong-gravity-regime" in the space of gravitational couplings.
In $d=4$, it is possible to remain in the allowed region while simultaneously solving the Landau-pole problem. As $d$ is increased, the viable area in the gravitational parameter space where this is possible without entering the ``strong-gravity regime", shrinks to zero within the explored range.

Thus, $d=4$ dimensions appears to be special in asymptotically safe matter-gravity models as the only dimensionality that can accommodate an ultraviolet complete Abelian sector. According to our results, $d=5$ might work, but only within a tiny part of the gravitational parameter space.\\
One might expect an embedding into a grand unified theory (GUT) to provide a way to circumvent this constraint. In $d=4$ dimensions, GUTs can be asymptotically free, or feature Landau poles, depending on the matter content, see, e.g., \cite{Bertolini:2009qj}. The former become asymptotically safe in $d=4+\epsilon$ dimensions \cite{Peskin:1980ay,Gies:2003ic,Morris:2004mg} even in the absence of gravity. 
Yet, even without gravity there is an upper critical dimension beyond which asymptotic safety in non-Abelian SU($N$) gauge theories is lost \cite{Gies:2003ic} and these theories are no longer UV complete on their own. 
With gravity, we expect this upper critical dimension to be shifted towards larger $d$, as the gravitational contribution $\sim f_g$ again acts akin to a dimensional reduction. However, this requires a strong growth of the Newton coupling as a function of $d$ to make the effect strong enough to induce asymptotic freedom.
On the other hand, the existence of an excluded strong-gravity region in parameter space is expected to persist in the non-Abelian case. In summary, we expect that embedding the Standard Model into a GUT could provide a way to arrive at a UV-complete model in $d=5$, but presumably not far beyond.

This indicates that the predictive power of the asymptotic-safety paradigm could extend to parameters beyond those characterizing the dynamics, and potentially even fix fundamental parameters of the geometry of spacetime. It is indeed remarkable that requiring a simple description of nature which does not need degrees of freedom beyond those that have already been observed, might work at all -- establishing whether it does of course requires to check additional conditions beyond the UV completion of the Abelian sector -- and furthermore actually appears to single out $d=4$ as the only potentially
 viable choice of dimensionality.

It is a particularly intriguing question whether a similar argument pertains to the Lorentzian case of $3+1$ dimensions.

\emph{Acknowledgements:}
This work is supported by the DFG under grant no.~Ei-1037/1.

\bibliography{refs}
\end{document}